# Charge redistribution at metal–$ZrO_2$ interfaces: A combined DFT and continuum electrostatic study

Ximeng Wang*[1], Yongfeng Zhang*[1], Dmitry Skachkov[2], Arnab Das[1,3], Junliang Liu[4], Alexander Kvit[1,6], Jennifer Choy[5], and Adrien Couet[1]

[1]Department of Nuclear Engineering and Engineering Physics, University of Wisconsin-Madison, 1500 Engineering Dr, Madison, WI 53706, USA

[2]NanoScience Technology Center and Department of Physics, University of Central Florida, Orlando, FL 32826, USA

[3]Department of Materials Science and Engineering, University of Wisconsin-Madison, Materials Science and Engineering Building, 1509 University Avenue Madison, WI 53706, USA

[4] Department of Materials Science and Engineering, FAMU–FSU College of Engineering, Florida State University, 2525 Pottsdamer St., Tallahassee, FL 32310, USA

[5]Department of Electrical and Computer Engineering, University of Wisconsin-Madison, Engineering Hall, 1415 Engineering Drive, Madison, WI 53706, USA

[6]The Nanoscale Imaging and Analysis Center, University of Wisconsin-Madison, Materials Science and Engineering Building 1509 University Avenue Madison, WI 53706, USA

* Corresponding authors: xwang2782@wisc.edu, yzhang2446@wisc.edu

**Abstract**

Nanoscale metallic inclusions (NMIs) are commonly observed within oxide scales formed during high-temperature oxidation, revealing the existence of chemical and electronic heterogeneity beyond conventional corrosion theories that assume homogeneous, fully oxidized films. Using tetragonal zirconia ($tZrO_2$) facing a series of face-centered cubic (fcc) metals as the model system, this work investigates the short-range and long-range charge redistributions across metal-oxide interfaces by coupling density functional theory (DFT) calculations with continuum modeling. We show that metal-oxide contact induces a short-range charge redistribution confined to a few atomic layers and a long-range redistribution of space charge that can extend over macroscopic distances within weakly doped oxides. DFT calculations show that the short-range redistribution is dominated by metal induced gap states (MIGS) in $tZrO_2$ facing noble metals like Au and Ag, and by chemical bonding in $tZrO_2$ facing active metals like Al. DFT-informed continuum theoretical analysis shows that the range of space-charge redistribution is governed by the doping level of $tZrO_2$, and that the Schottky barrier height (SBH) exhibits a stronger dependence on the metal work function than the doping level. Both the short-range and long-range charge redistributions can alter the transport of charge carriers via their associated electric fields, extending several nm to hundreds of nm from the interface, depending on the doping concentrations, suggesting possible heterogeneous oxide growth caused by NMIs.

## 1. Introduction

Corrosion is a general degradation phenomenon in metals and alloys. Understanding the corrosion mechanism is critical for designing corrosion-resistant alloys. The traditional mechanism assumes that oxide layers are fully homogeneous in terms of both chemistry and microstructure, implying homogeneous oxide growth.[1–3] This widely accepted assumption of oxide homogeneity is often violated in alloys subjected to high-temperature oxidation, where nanoscale heterogeneities, either chemically or structurally or both, have been commonly observed in growing oxides, invalidating the homogeneous oxide assumption. For example, in nuclear reactor environments, nanoscale precipitates undergo progressive oxidation as the protective oxide layer grows on fuel cladding materials. These precipitates have been widely observed in irradiated and unirradiated n-type oxides, such as metallic or oxidized $Zr(Fe,Cr)_2$ Laves phases in $ZrO_2$ grown on Zr-based nuclear fuel cladding containing Fe and Cr, or metallic or oxidized βNb particles in $ZrO_2$ grown on ZrNb alloys fuel cladding. [4–8] Such nanoscale metallic inclusions (NMIs) embedded in the oxide phase introduce internal metal–oxide interfaces that perturb local electrostatic potentials and electric fields, which induce heterogeneities in the fluxes of charged species (e.g., electrons and oxygen vacancies) towards the growing oxide surfaces, potentially causing inhomogeneous oxide growth. Consistent with this picture, non-planar and locally accelerated oxide growth has been frequently reported in the presence of embedded metallic or semi-metallic phases [5,9,10].

Elucidating the role of NMIs in oxide growth requires a mechanistic understanding of how they influence the transport of charged species, which in turn depends on how NMIs modify the electrostatic fields within the oxide. Analogous to metal–semiconductor interfaces [11], the formation of a metal–oxide interface induces both short-range and long-range charge redistributions, accompanied by the formation of a Schottky barrier and an electric field that affect charged-species transport [12]. When a metal is brought into contact with an oxide (or a semiconductor), charge rearrangement occurs at the interface due to quantum-mechanical effects such as wavefunction overlap, hybridization, metal-induced gap states (MIGS), and chemical bonding. These effects give rise to an interface dipole and a corresponding shift in the electrostatic potential. This short-range contribution is confined to one or two screening lengths in the metal and a few atomic layers on the oxide side, as consistently demonstrated by density functional theory (DFT) calculations. [13–19]. The magnitude and direction of this interfacial charge redistribution depend on the oxide's defect chemistry and doping level [16,19–21], as well as on the surface orientation and termination [17,22].

Beyond this short-range redistribution localized near the interface, the establishment of a metal–oxide contact can also drive a long-range redistribution of space charge within the oxide. This process is governed by Fermi-level alignment between the metal and the oxide, subject to possible Fermi-level pinning by interfacial states, and extends over a length scale set by the Debye screening length [23]. The resulting accumulation or depletion of charge carriers near the interface produces an additional electrostatic potential drop within the oxide, distinct from that associated with the interface dipole. Consequently, the electric field and the Schottky barrier height depend on both the short-range interfacial contribution and the long-range space-charge distribution. Because the latter can extend far beyond the spatial scales accessible to atomistic simulations, it is typically treated using continuum electrostatic models [24].

Despite the technological importance of oxide scales in corrosion, the theoretical framework underlying metal-oxide interfaces remains far less developed than that for metal-semiconductor interfaces. Oxides formed during high-temperature oxidation are often ionic, wide band-gap materials. As such, they are intrinsically insulating or only weakly doped, with low free-carrier concentrations and long screening lengths. As a result, their space-charge behavior can differ from that of conventional semiconductors. Furthermore, NMIs embedded within the oxide can vary substantially in chemical reactivity, work function, and valence electron configuration, each of which may influence interfacial charge redistribution in distinct ways. Disentangling the roles of these metal properties is therefore essential for understanding how different NMIs affect electrostatic fields and charged-species transport in oxides.

In the literature, short-range and long-range charge redistribution effects are commonly treated separately because of their disparate length scales. Short-range interfacial charge transfer is typically examined using DFT without consideration of space-charge effects, whereas long-range space-charge redistribution is addressed using electrostatic theories without explicit consideration of interface dipoles. As a consequence, the connection between atomistic charge redistribution and continuum space-charge behavior is yet to be fully established. In particular, interfacial charge transfer predicted by DFT is often on the order of 0.1-1 $e/nm^2$ [14,17,19,21,25–27], which can exceed continuum space-charge estimates by orders of magnitude, depending on the doping concentrations [28–30]. Without a unified framework to reconcile these descriptions, it remains unclear how short-range interfacial polarization and long-range space charge collectively shape electrostatic fields and transport processes in corrosion oxides.

In this work, we investigate charge redistribution at metal-oxide interfaces by combining first-principles DFT calculations with DFT-informed theoretical analysis. Tetragonal $ZrO_2$ ($tZrO_2$) facing a series of fcc metals with varying chemical activity, work function, and valence electron configuration is selected as the model system. This allows us to systematically disentangle electronic effects associated with metal-induced gap states from chemically driven charge transfer and bonding to apply a unified framework linking atomistic interfacial charge redistribution to long-range space-charge behavior in oxides. We demonstrate that metal-oxide contact induces both a short-range charge redistribution confined to a few atomic layers and a long-range space-charge redistribution that can extend over macroscopic distances within the oxide. Our DFT results show that the short-range redistribution is dominated by metal-induced gap states (MIGS) at interfaces with noble metals such as Au and Ag, whereas chemical bonding effects dominate at interfaces with active metals such as Al. The DFT-informed theoretical analysis further reveals that the spatial extent of space-charge redistribution is primarily governed by the oxide doping level, while the Schottky barrier height (SBH) exhibits a stronger dependence on the metal work function than on the oxide doping level.

## 2. Methodology

### 2.1 Materials system selection and interface model construction

In this work, we select $tZrO_2$ as a model ionic oxide representative of those formed during high-temperature oxidation. $ZrO_2$ is the primary corrosion product on Zr-based nuclear fuel cladding and plays a key role in limiting the safe operation of nuclear reactors [31,32]. Under normal operating conditions, the temperature remains below 1000 °C [33], and the monoclinic phase ($mZrO_2$) is thermodynamically the most stable phase [34,35]. However, the interface that forms between the metal and the $ZrO_2$ in the reactor is very complicated, which includes both monoclinic and tetragonal phases of zirconium oxide. Herein, we select $tZrO_2$, a commonly observed phase near the metal-$ZrO_2$ interface [36–40], likely stabilized by higher oxide compressive stresses, an increase in oxygen vacancy concentrations, or relatively significant aliovalent ion concentrations. While Fe and Cr are common dopants in Zr alloys, they have similar properties in valence electrons, work functions, and chemical reactivity, making it difficult to isolate the effect of individual metal properties on the oxide interface. To systematically explore the effect of metal properties on the oxide interface, three fcc metals - Au, Ag, and Al - are selected, as they offer a wider range of work functions, chemical reactivities, and valence electron configurations. As shown in Figure 1, the work functions of the {111} surfaces of Au, Ag, and Al range from 5.15 eV to 4.02 eV below the vacuum level. For comparison, $tZrO_2$ with a (100) surface has a work function of 5.80 eV, an electron affinity of 1.18 eV, and its Fermi level is located near the middle of the band gap at 3.49 eV, as obtained from our DFT calculations (Table 1). Au and Ag, as noble metals, are expected to exhibit minimal bonding with oxygen (O) in $ZrO_2$, whereas Al, an active base metal, is expected to form bonds with oxygen. Furthermore, the 5d and 4d valence electrons of Au and Ag differ significantly in their orbital character from the 3s and 3p valence electrons of Al, leading to distinct coupling behaviors with O 2p and Zr 4d orbitals because of orbital symmetry and energy levels. Although these metals are not typically present in nuclear fuel cladding, they provide an ideal platform to systematically assess the roles of chemical reactivity, work function, and valence electron configuration on metal-oxide interfacial behavior.

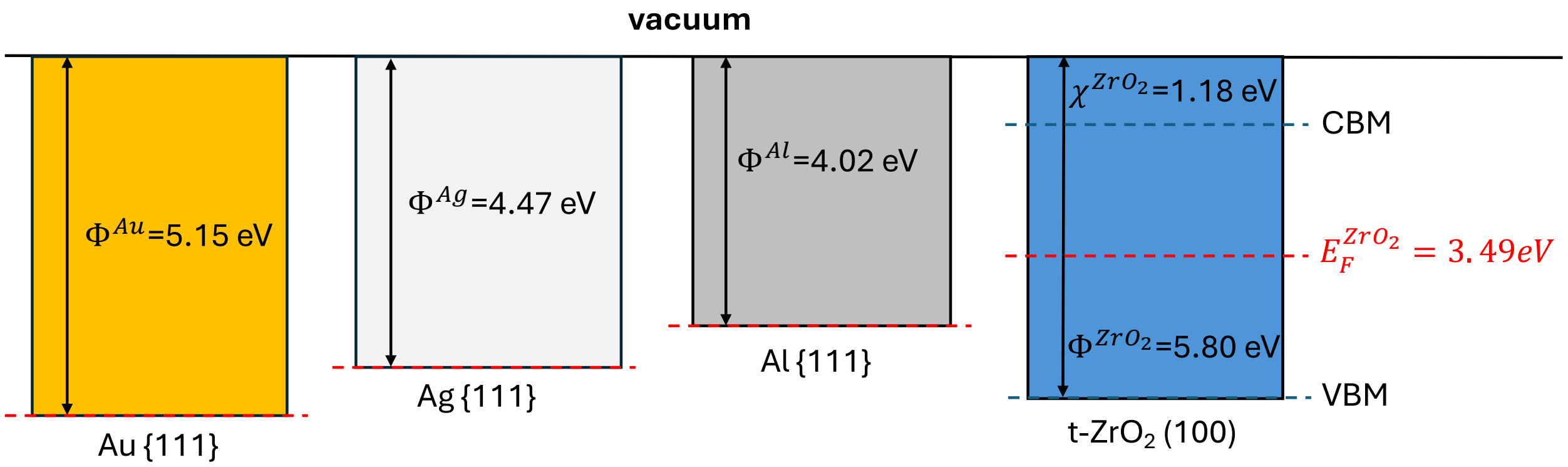

Figure 1. Schematic showing the metal and $ZrO_2$ work functions ($\Phi$) and Fermi level ($E_F$, dashed red line), and the electron affinity ($\chi$) of $tZrO_2$. The values are from the present DFT calculations.

To construct the interface structures, low-index planes are selected for both the metal and the oxide. For Au, Ag, and Al, the (111) surface is chosen because it has the lowest surface energy [41]. For $tZrO_2$, one of its low-index surfaces, (100), is selected to form interfaces with the metals. [42,43] For convenience, the $z$ axis is defined along the [111] direction of the metals and the [100] direction of $tZrO_2$. The $x$ and $y$ axes are chosen as $[11\bar{2}]$ and $[1\bar{1}0]$ for the metals, and [010] and [001] for $tZrO_2$, respectively, as shown in Figure 2. This orientation relationship is determined by minimizing the lattice mismatch strain while aligning the densely packed directions of the fcc metals and $tZrO_2$. The metal region contains 6 (111) layers, and the oxide region contains 12 (100) layers, balancing computation cost and accuracy. The cell dimensions along $x$ and $y$ are 0.52 and 1.47 nm, respectively. The orientation-dependent mismatch strains obtained with the data from the Materials Project [44] are summarized in Table 1. Previous work on Au–$TiO_2$ [45]and graphene–$MoS_2$ [46]interfaces has shown that 1–5% strain can lead to a 0.1–0.5 eV change in the Schottky barrier height (SBH). Based on these literature results, the uncertainty in SBH induced by the mismatch strain, which is within 5% in this work, is estimated to be within approximately 0.5 eV.

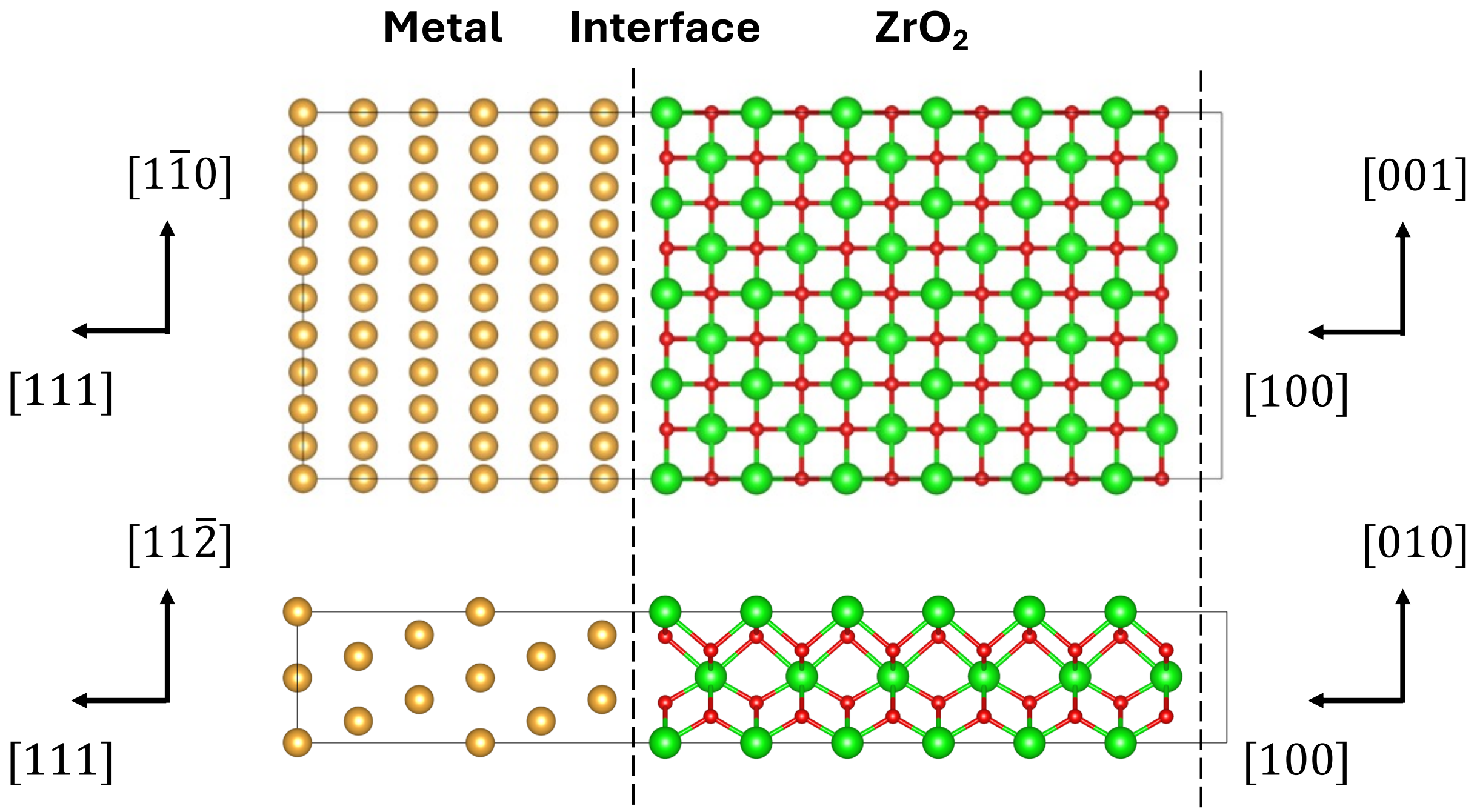

Figure 2. The metal (111)/$tZrO_2$ (100) interface and the orientation relationship viewed from (top) the [010] and (bottom) [001] orientations of $tZrO_2$.

Table I. Metal work functions and lattice mismatch in % along x and y axes taking metals as the references. Positive (negative) values indicate the $ZrO_2$ lattice parameter is larger (smaller). The work function data from the literature are shown in parentheses.

| Metal | (111) work function (eV) | x mismatch with $ZrO_2$ (%) | y mismatch with $ZrO_2$ (%) |
|---|---|---|---|
| Au | 5.15 (5.11-5.22[47–49]) | 2.35 | -2.37 |
| Ag | 4.47 (4.36-4.74[48,50,51]) | 3.98 | -0.76 |
| Al | 4.02 (4.02-4.21[47,52,53]) | 5.45 | 0.84 |

### 2.2 DFT settings

All properties of metal-t$ZrO_2$ interface structures are calculated using DFT as implemented in the Vienna *ab initio* Simulation Package (VASP). [54,55] The projector augmented-wave (PAW) method is employed to describe the electron-ion interactions.[56,57]. Exchange-correlation effects are treated within the generalized gradient approximation (GGA) using the Perdew-Burke-Ernzerhof (PBE) functional. [58]. The valence electrons included in the calculations are Al $3s^2 3p^1$, Au $5d^{10} 6s^1$ and Ag $4d^{10} 5s^2$, Zr $4s^2 4p^6 4d^3 5s^1$, and O $2s^2 2p^4$, corresponding to 3, 11, 12, 12, and 6 valence electrons, respectively. A plane-wave cutoff energy of 500 eV is used for all calculations. Monkhorst–Pack $k$-point meshes of $4 \times 4 \times 4$ and 12 x 12 x 4 are employed for structural relaxation and electronic structure calculations, respectively. [59]. Structural relaxations are performed until the maximum residual force on each atom is less than 0.02 eV/Å. To ensure numerical stability and consistent force convergence in the large interface supercell, Gaussian smearing was employed with a width of 0.05 eV.

Spin polarization is included in all calculations. To improve the description of the localized Zr 4d states, the DFT+U method is applied to Zr atoms with an effective $U - J = 4.0$eV [60-62], which gives a reasonable band gap value of 4.615 eV, in comparison to experimentally measured values ranging from 4.2 to 5.7 eV [63–65]. All structures are relaxed at fixed cell volumes. All the atoms are allowed to move during relaxation. Structural visualization is carried out using the VESTA software [66]. Bader charge analysis is conducted using the code developed by the Henkelman Group [67,68].

### 2.3 Work function and energy of adhesion

To understand band alignment and charge redistribution at the metal-t$ZrO_2$ interface, the work functions of the Au(111), Ag(111), and Al(111) surfaces, as well as the t$ZrO_2$ (100) surface, are calculated. To ensure a reliable representation of both surface and bulk properties, the fcc metal slabs consist of nine atomic layers, while the t$ZrO_2$ slab contains sixteen layers. Each slab model includes a 20 Å vacuum region to accurately determine the vacuum level. The work function is defined as the energy difference between the vacuum level and the Fermi level ($E_F$) for the metals, or between the vacuum level and the valence band maximum (VBM) for $tZrO_2$, as illustrated in Figure 1. The results are listed in Table 1. The good agreement between our calculations and the literature data justifies the DFT settings.

The adhesion energy between the fcc metals and $tZrO_2$ is also computed following the approach of Muñoz *et al.* [18]. It is defined as the energy difference per unit interfacial area between the combined metal-oxide system with the interface (see Figure 2) and a reference state consisting of the corresponding metal and oxide slabs separated without interfacial contact.

### 2.4 DFT-informed calculations of long-range space charge redistribution

Due to the limited supercell size accessible in periodic DFT calculations, long-range redistribution of space charge cannot be explicitly captured. This contribution is instead evaluated by solving Poisson's equation in the oxide region using the self-consistent algorithm developed by Skachkov *et al.* [24,69], which is well suited for incorporating DFT-derived interfacial electronic properties into a continuum electrostatic framework.

In this approach, the metal is treated as an ideal electron reservoir with perfect screening, so that all net charges accumulate at the interface. The metal-oxide interface is characterized by an interfacial potential offset, while the electrostatic potential far from the interface in the oxide is taken as the reference and set to zero. Only the redistribution of space charge in the oxide driven by Fermi-level equilibration is considered. The oxide is modeled as a lightly doped ionic solid with a low bulk space-charge density (i.e., free electrons or holes), $\rho_0$, which becomes either enriched or depleted near the interface in response to band bending induced by the electrostatic potential. This band bending shifts the band edges and thereby modifies the occupation of bulk electronic states at a given temperature, changing the local space charge density.

Conversely, the electrostatic potential, which is induced by dielectric polarization and space charge redistribution, is determined from the space charge density profile by solving Poisson's equation. The boundary condition at the metal-oxide interface is set by the net interfacial charge density, which is computed by integrating the DFT-derived interfacial states between the charge neutrality level (CNL) and the Fermi level, and polarization at the interface. In the bulk oxide, charge neutrality is enforced by taking the potential far from the interface as the reference state. Global charge neutrality further requires that the net interfacial charge be balanced by the total change in space charge within the oxide.

Poisson's equation is therefore solved iteratively in a self-consistent manner until both Fermi-level equilibrium and global charge neutrality are simultaneously satisfied. Details of the numerical implementation and convergence criteria can be found in the original work of Skachkov *et al.* [24]. The converged space charge density and electrostatic potential profiles are then used to determine the SBH, depletion layer width (DLW), and net charge transfer across the interface (i.e., the interfacial charge density). Specifically, the DLW is defined as the distance from the interface at which the magnitude of the space-charge density decays to $1/e$ of its interfacial value.

## 3. Results and discussion

### 3.1 Interfacial structure and chemistry

The interfaces between the (111) surfaces of Au, Ag, and Al and the (100) surface of $tZrO_2$ exhibit relatively weak adhesion, with adhesion energies of 0.29, 0.16, and 0.60 $J/m^2$, respectively, as listed in Table II. The results are consistent with the 0.156 $J/m^2$ adhesion energy reported for the interface between Ag (111) and cubic $ZrO_2$ (111) [19]. In contrast, significantly stronger adhesion has been predicted by DFT calculations for interfaces such as Ni(001)-c-$ZrO_2$(001) [18], Pd(111)-ZnO $(000\bar{1})$ [13], and Ti(0001)-$TiO_2$(110) [17], indicating stronger interfacial chemical interactions in these systems.

Figure 3 presents the relaxed atomic structures of the metal-$tZrO_2$ interfaces (top panels) together with the corresponding differential electron density maps (bottom panels). The differential electron density, $\Delta\rho(r)$, is obtained by subtracting the superposed electron densities of the isolated metal and oxide slabs from that of the combined interface system, while keeping the ionic positions fixed to those of the relaxed interface configuration. This procedure isolates charge redistribution arising purely from interfacial interactions rather than structural relaxation effects.

For the Au-$tZrO_2$ and Ag-$tZrO_2$ interfaces shown in Figures 3(a) and 3(b), no significant atomic rearrangements are observed near the interface. The metal regions retain their fcc structure, and the oxide regions preserve the tetragonal structure of $tZrO_2$. Only minor interfacial relaxations are present, with some surface metal atoms moving towards the oxide. Importantly, the differential electron density maps show negligible charge accumulation in the interfacial region. In fact, electron depletion is observed between the closest Au/Ag-O atomic pairs (see the dotted squares in the bottom panels), indicating the absence of directional bonding. This behavior is characteristic of noble metal-oxide interfaces and is consistent with the chemically inert nature of Au and Ag.

Similar weak charge redistribution and lack of covalent bonding have been widely reported for noble-metal/oxide interfaces, including Pt(111)-$CeO_2$(111) [70], Pt(001)-$TiO_2$(001) [71], Au-$\alpha Fe_2O_3$(001) [72], Ag(001)-MgO(100) [73], and Ag(111)-$ZrO_2$(111) [19], in both experimental and first-principles studies.

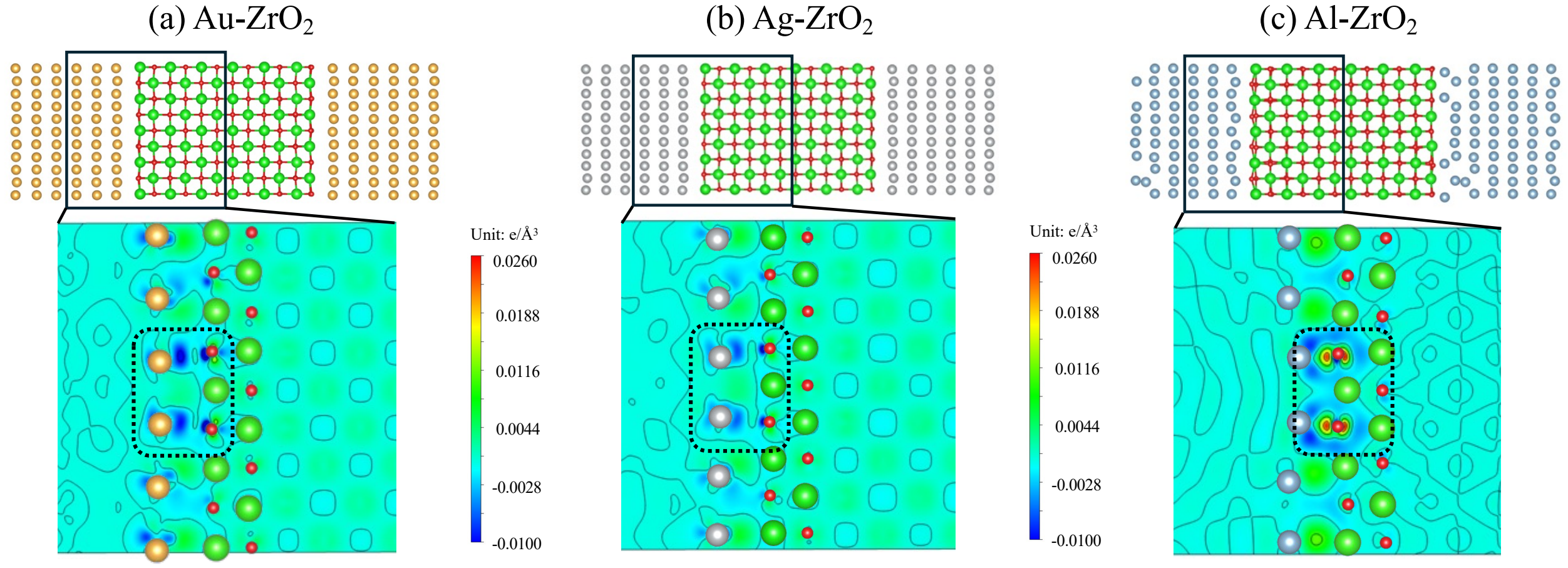

Figure 3. (Top) relaxed interface structures viewed from the [010] axis of $tZrO_2$ and (bottom) differential electron density, $\Delta\rho(r)$, across the interface for (a) Au-$ZrO_2$ (b) Ag-$ZrO_2$ (c) Al-$ZrO_2$ interfaces. Periodic images of the metal regions are added for visualizations. Au, Ag, and Al atoms are colored in yellow, silver, and light blue, and Zr and O in green and red, respectively.

In contrast, the Al-$tZrO_2$ interface shown in Figure 3(c) exhibits much stronger structural relaxation and electronic reorganization. Two distinct interfaces are present in the simulation cell. At the left interface, atomic displacements are qualitatively similar to those observed for Au and Ag but are larger in magnitude. At the right interface, the Al atoms reorganize into two atomic layers, likely driven by the larger lattice mismatch strain and the stronger chemical affinity between Al and O. Correspondingly, the differential electron density map at the left interface reveals pronounced electron accumulation between neighboring Al and O atoms (see the dotted square in the bottom panel), clearly indicating the formation of interfacial Al-O bonds. This behavior reflects the high chemical reactivity of Al and its strong tendency to form oxides, in contrast to Au and Ag. Comparable strong charge transfer and bonding have been reported for reactive metal-oxide interfaces such as Ni(111)-$\alpha Al_2O_3$(0001) [20,74], Ni(111)-$ZrO_2$(111) [74], Al(111)-$\alpha Al_2O_3$(0001) [75,76] where significant covalent or partially ionic bonding dominates the interfacial interaction. We have also observed the same trend in the differential electron density map at the right interface. The details are shown in Figure S1 in the Supplementary Materials.

Overall, the results in Figure 3 highlight a clear transition from weak interfaces for noble metals to chemically bonded interfaces for active metals. This distinction is critical for understanding the different regimes of short-range charge transfer that will be discussed in the following section.

Table II. Adhesion energy, charge transfer, and electrostatic potential drop obtained from DFT calculations.

| **Interface** | **Adhesion energy ($J/m^2$)** | **Charge transfer (e/nm2)** | **Potential drop (V)** |
|---|---|---|---|
| Au-$ZrO_2$ | 0.29 | -0.45 | 0.49 |
| Ag-$ZrO_2$ | 0.16 | -0.03 | -0.74 |
| Al-$ZrO_2$ | 0.60 | 2.15 | -2.54 |

### 3.2 Short-range charge redistribution at the interface

In Figure 4(a)-(c), the layer-averaged Bader charges are plotted along the interface normal for the Au-t$ZrO_2$, Ag-t$ZrO_2$, and Al-t$ZrO_2$ interfaces. For each atomic layer, the Bader charges of all atoms are summed and then normalized by the number of atoms in the corresponding layer of the unrelaxed structure, allowing a direct comparison of charge redistribution across the interface.

Despite the minimal structural relaxation observed for the Au-t$ZrO_2$ and Ag-t$ZrO_2$ interfaces, pronounced electronic charge redistribution occurs in all three systems. For the Au-t$ZrO_2$ interface (Fig. 4(a)), the first Au layer develops a net negative charge upon contact, while the adjacent $ZrO_2$ layer becomes positively charged. This behavior closely resembles the charge redistribution reported for the Au–$TiO_2$ interface by Jiao *et al.* [16], where electron transfer driven by Fermi-level alignment produces an interfacial dipole without significant chemical bonding. The slight positive charging of deeper Au layers is likely a finite-size effect associated with the limited number of metal layers in the DFT supercell, rather than a physical depletion region in the metal. On the oxide side, the positive charge decays within approximately four $ZrO_2$ layers, beyond which charge neutrality is recovered, indicating that the charge redistribution is confined to a few atomic layers near the interface.

A qualitatively similar behavior is observed for the Ag-t$ZrO_2$ interface (Fig. 4(b)), although in this case the first $ZrO_2$ layer is also slightly negatively charged. The overall magnitude of charge redistribution is smaller than in the Au case.

In contrast, the Al-t$ZrO_2$ interface exhibits markedly different behavior (Fig. 4(c)). Here, the first Al layer becomes positively charged, while the adjacent $ZrO_2$ layer is negatively charged, reflecting substantial electron transfer from Al to $ZrO_2$. This charge redistribution correlates with the formation of Al-O chemical bonds identified in the differential charge-density maps and structural relaxations. Notably, the charge perturbation in $ZrO_2$ is highly localized, with all subsurface oxide layers remaining essentially charge neutral. This indicates that the interfacial charge transfer in Al-t$ZrO_2$ is dominated by short-range chemical bonding, in contrast to the noble-metal interfaces.

The net electron transfer across each interface is summarized in Figure 4(d), where positive values denote electron flow from the metal into t$ZrO_2$. For the Au-t$ZrO_2$ interface, electrons transfer from $ZrO_2$ to Au, with a magnitude of 0.45 e/nm$^{-2}$ (Table II), comparable to the charge transfer reported for Au-$TiO_2$ by Jiao *et al.* [16]. The Ag-t$ZrO_2$ interface shows a much smaller charge transfer (-0.03 e nm$^{-2}$), indicating near-neutral behavior. In sharp contrast, the Al-t$ZrO_2$ interface exhibits a large electron transfer from Al to $ZrO_2$ of 2.15 e/nm$^{-2}$. The significantly larger magnitude and opposite sign relative to the noble-metal interfaces underscore the dominant role of chemical bonding, consistent with previous studies of reactive metal–oxide interfaces such as Al-$Al_2O_3$ [77] and Ti-$TiO_2$ [17]. As discussed by Puigdollers et al.[78], when a chemically, low work function metal contacts an oxide with high electron affinity, electrons can transfer directly from the metal to the oxide, enriching the oxide surface in electrons.

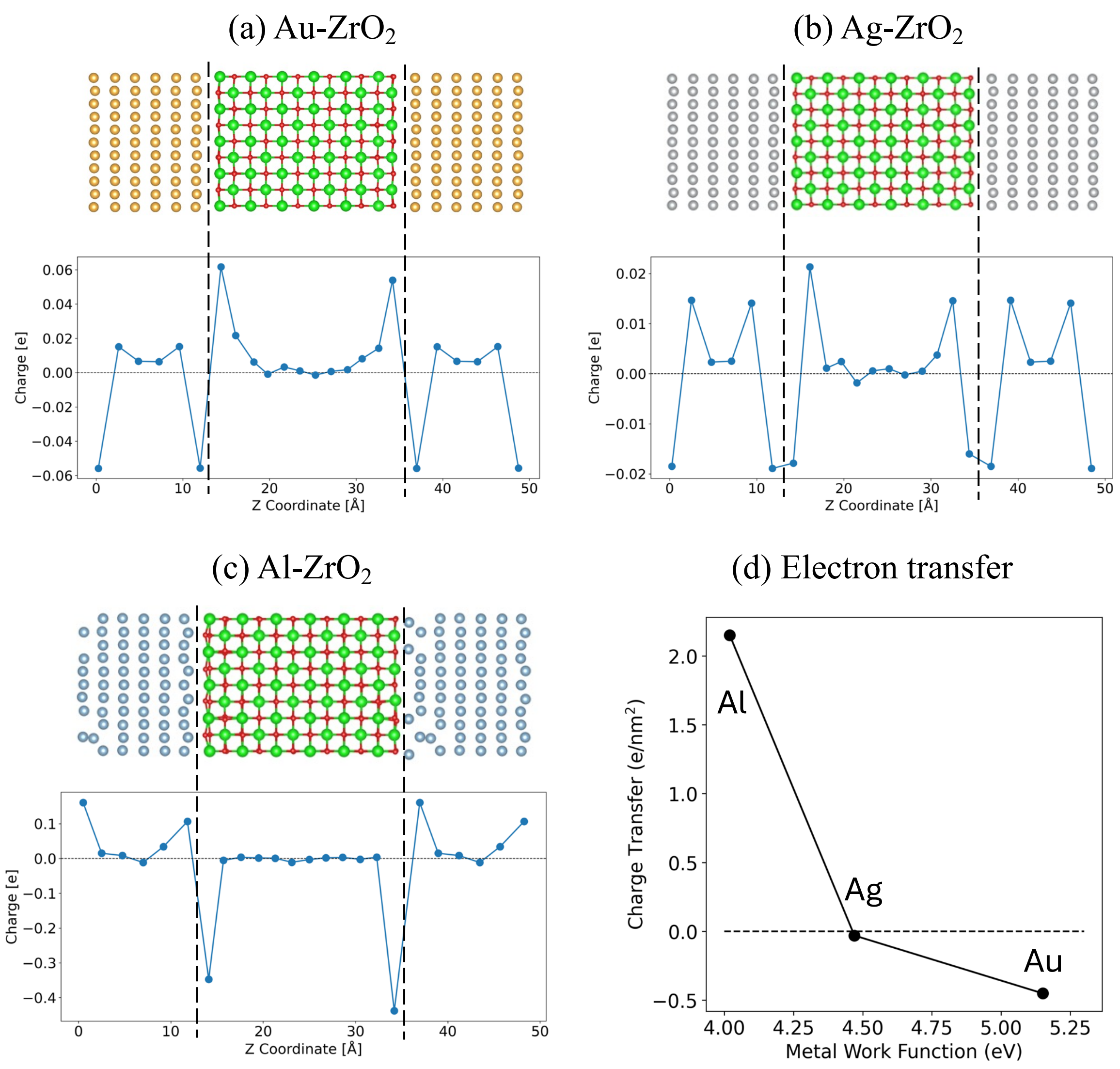

Figure 4. (in a-c, Top) relaxed interface structures viewed from the [010] axis of $tZrO_2$ and (bottom) normalized Bader charge in each layer per atom for (a) Au-$ZrO_2$ (b) Ag-$ZrO_2$ (c) Al-$ZrO_2$ interfaces (marked by the vertical dashed lines). In (d), the total electron transferred from metals to $tZrO_2$ is plotted against their work functions. Positive means electrons move from metal to $tZrO_2$.

The interfacial charge redistribution is expected to generate an electrostatic dipole and corresponding potential step across the interface. To elucidate this effect, electrostatic potential profiles obtained from DFT are shown in Figure 5(a)-(c). To suppress rapid oscillations arising from the ionic cores and localized electron density, the raw potentials are macroscopically averaged. In all cases, the averaged potentials reach well-defined plateaus in the bulk metal and bulk oxide regions, connected by a narrow transition region spanning only a few atomic layers.

This behavior confirms that the interfacial dipole is highly localized, in agreement with the spatial extent of the Bader charge redistribution.

The resulting potential drops between the bulk oxide and bulk metal regions are plotted in Figure 5(d) and listed in Table II, where a positive value corresponds to a higher electrostatic potential in the oxide. For the Au-t$ZrO_2$ interface, the potential drop is approximately 0.49 V, consistent in both magnitude and sign with electron transfer from $ZrO_2$ to Au. This charge transfer leaves the oxide region positively charged and establishes a built-in electric field that opposes further electron transfer. The net positive charge in the oxide region leads to an elevated electrostatic potential relative to the metal.

For the Ag-t$ZrO_2$ interface, the averaged potential in Ag is approximately 0.74 V higher than in bulk $ZrO_2$, yielding a sign that appears inconsistent with the small net electron transfer inferred from Bader analysis. Given the extremely small magnitude of charge transfer in this system, this discrepancy is likely attributable to the known limitations of Bader partitioning at interfaces, where fixed atomic basins may inadequately capture subtle charge rearrangements. In such cases, the sign of the extracted charge transfer can be sensitive to the analysis method, while the electrostatic potential profile provides a more robust indicator of the interfacial dipole.

For the Al-t$ZrO_2$ interface, the potential in bulk Al is 2.54 V lower than that in bulk $ZrO_2$, fully consistent with the large electron transfer from Al to the oxide and the formation of a strong interfacial dipole. The magnitude of this potential step further reflects the chemically reactive nature of the interface, in line with prior first-principles studies of reactive metal-oxide interfaces [14].

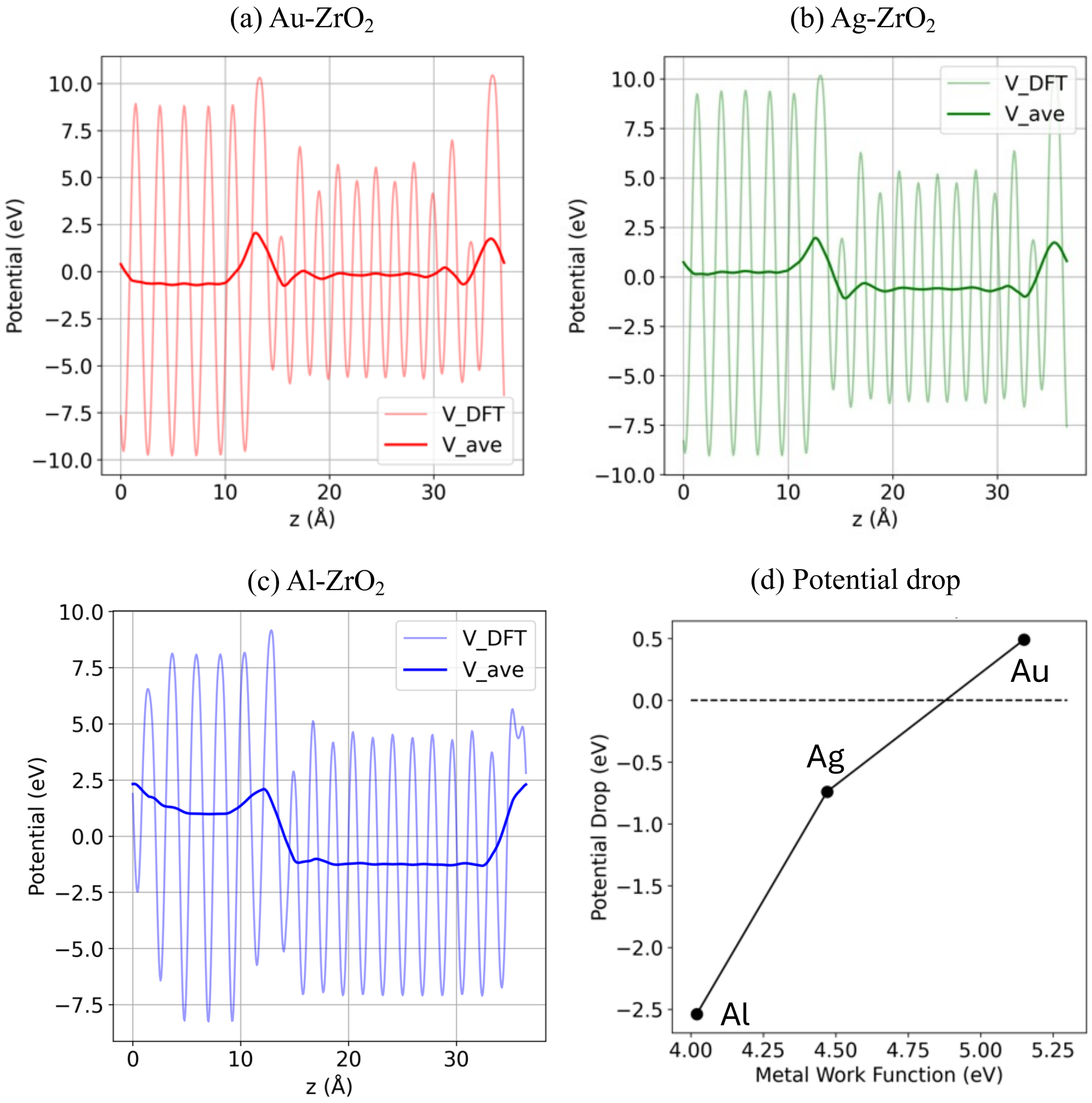

Figure 5. Raw DFT (thin curves) and averaged (thick curves) electrostatic potential profiles obtained from DFT across the (a) Au-t$ZrO_2$ (b) Ag-t$ZrO_2$ (c) Al-t$ZrO_2$ interfaces. In (d), the drops in potential, i.e., the averaged potentials in the bulk t$ZrO_2$ in reference to those in bulk metals, are plotted against the metal work function.

### 3.3 Metal induced gap state and Schottky barrier height

The Bader charge analysis and electrostatic potential profiles presented in the previous section establish that metal-t$ZrO_2$ contact universally induces a localized interfacial dipole, whose magnitude and sign depend sensitively on the metal species. For noble metals, this dipole arises in the absence of chemical bonding and is therefore electronic in origin, suggesting a key role of MIGS and Fermi-level alignment. In contrast, at the Al-t$ZrO_2$ interface, strong chemical bonding leads to substantial charge transfer and a much larger interfacial dipole dominated by localized Al-O hybridization. To elucidate the microscopic origin of these differences and to clarify how metal properties control the short-range charge redistribution, we now examine the nature and spatial decay of MIGS in the oxide.

Figure 6 shows the layer-resolved projected density of states (PDOS) for $ZrO_2$ in contact with Au, Ag, and Al. Only O 2p and Zr 4d states are shown, as other O and Zr orbitals contribute negligibly to the gap states. Because the finite size of the interface supercells can shift the band edges relative to bulk $ZrO_2$, the DOS spectra of the interface cells are aligned to bulk by matching the deep core-like states (O 2s and Zr 4s/4p), as illustrated in Figure S2 of the Supplementary Materials. After this alignment, the PDOS of the sixth oxide layer from the interface closely resemble that of bulk $tZrO_2$ (Figure S3 of Supplementary Materials), confirming that the 12-layer oxide slabs are sufficient to recover bulk-like behavior away from the interface.

As shown in Figure 6(a) and (b), contact with Au and Ag induces substantial MIGS within the band gap of $ZrO_2$. Near the VBM, the MIGS are dominated by O 2p character. In particular, a pronounced peak appears just above the VBM in the first oxide layer, originating from the hybridization of interfacial O 2p states that lie below the VBM in bulk $ZrO_2$. These O-derived MIGS decay rapidly into the oxide and become negligible beyond approximately three oxide layers, as highlighted by the insets in the upper panels of Figures 6(a) and (b). Near the CBM, the MIGS are primarily composed of Zr 4d states and exhibit a similarly short decay length, again vanishing within roughly three layers (see insets in the bottom panels of Figures 6(a) and (b)). This spatial confinement is consistent with the canonical picture of MIGS at metal-insulator interfaces, where evanescent metal states penetrate only a few atomic layers into the wide-gap oxide.

To quantify the decay behavior, Figure 7 plots the total integrated DOS within the band gap as a function of oxide layer index. The Au-$tZrO_2$ and Ag-$tZrO_2$ interfaces exhibit very similar MIGS magnitudes and decay trends. Within the first three oxide layers, the total MIGS density decreases by nearly two orders of magnitude (note the logarithmic scale of the y-axis). While MIGS are often assumed to decay exponentially, the slight deviation from a perfect exponential form observed here likely reflects the finite thickness of the oxide region, which limits the ability to fully capture the asymptotic decay regime for the Au-$tZrO_2$ and Ag-$tZrO_2$ interfaces.

In contrast, the Al-$tZrO_2$ interface exhibits fewer MIGS, and a significantly faster decay, as shown in Figure 6(c) and Figure 7. As in the noble-metal cases, the MIGS near the VBM are dominated by O 2p states and those near the CBM by Zr 4d states. However, the overall density of gap states is lower, and the decay length is shorter. Notably, there is no O 2p-derived peak within the gap and an enhanced Zr 4d contribution near the CBM compared to the Au and Ag cases. As a consequence, the Fermi level in the Al-t$ZrO_2$ interface cell is pinned much closer to the CBM, at approximately 3.00 eV above the VBM, compared to 1.48 eV for Au-t$ZrO_2$ and 2.02 eV for Ag-t$ZrO_2$.

The reduced density and faster decay of MIGS at the Al-t$ZrO_2$ interface can be traced to fundamental differences in the valence electron configurations of the metals. Unlike the spatially extended 4d and 5d states of Ag and Au, the Al 3s and 3p valence states have different symmetry and energy alignment relative to O 2p orbitals, resulting in weaker evanescent coupling across the interface. Consequently, MIGS play a secondary role in determining the interfacial electronic structure for Al-t$ZrO_2$. Instead, the large charge transfer observed for Al is dominated by direct chemical bonding and orbital hybridization, consistent with the strong Al-O interactions identified in the charge-density and structural analyses.

(a) Au-$ZrO_2$

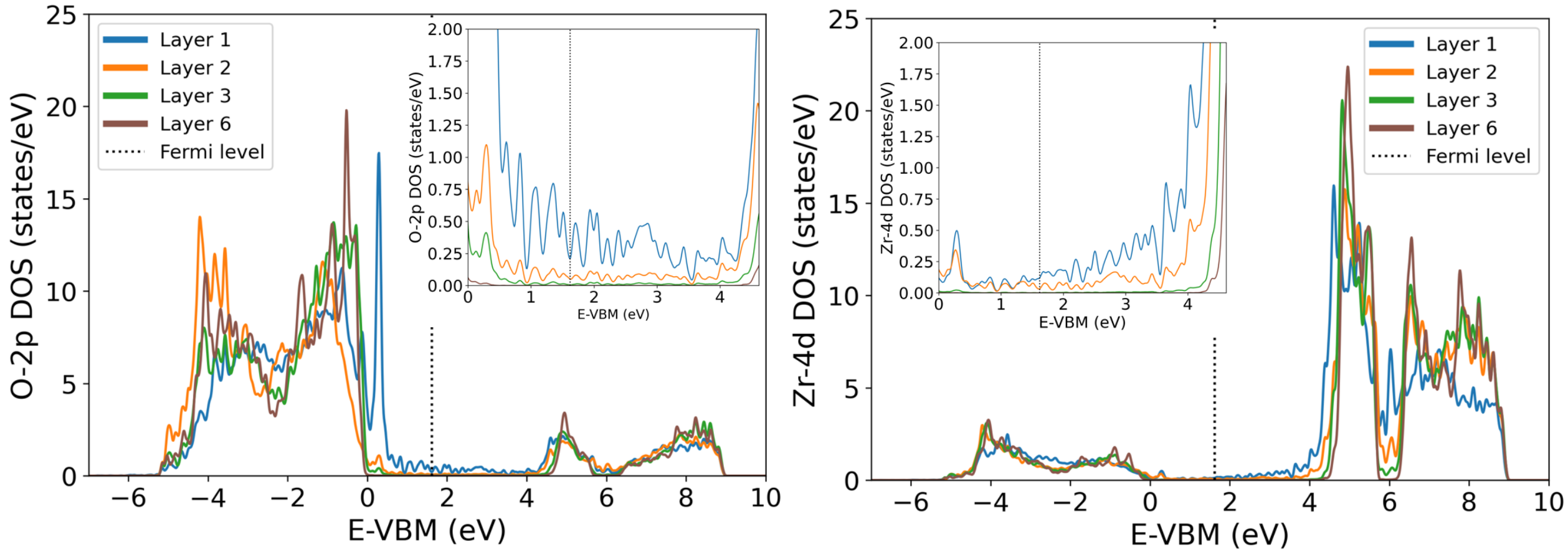

(b) Ag-$ZrO_2$

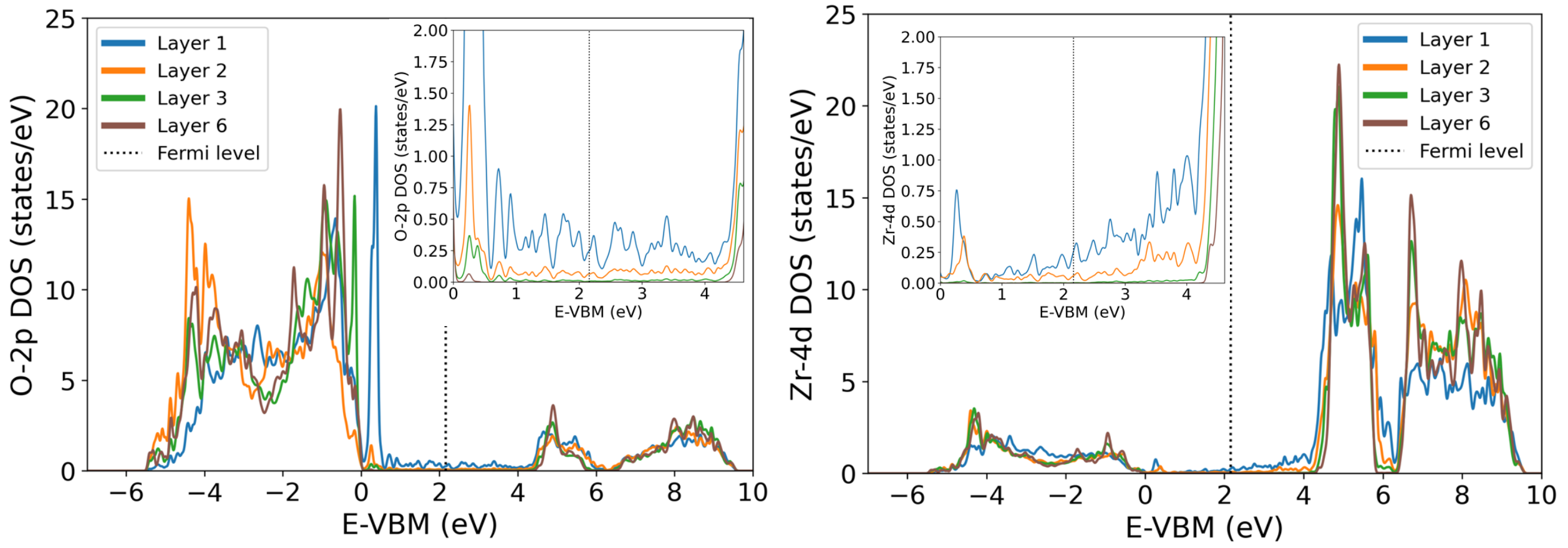

(c) Al-$ZrO_2$

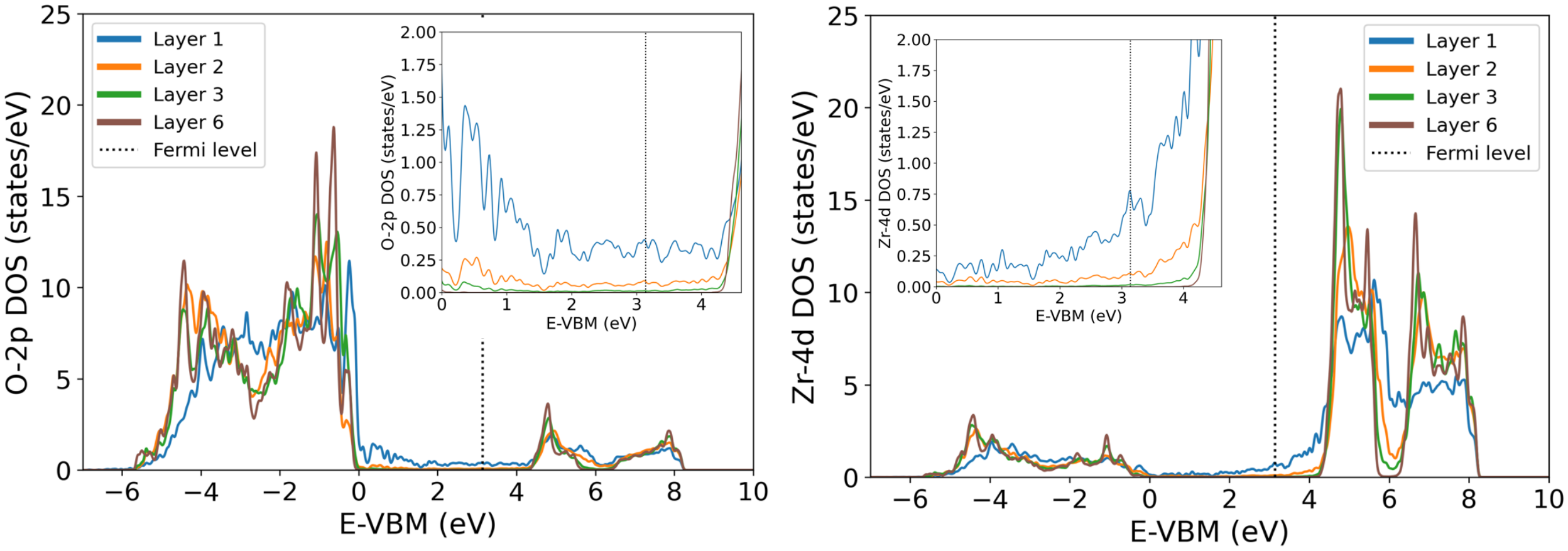

Figure 6. Projected density of states (PDOS) in $tZrO_2$ on contact with (a) Au, (b) Ag, and (c) Al.

The PDOS of the first three oxide layers are plotted to compare with the $6^{th}$ oxide layer representing bulk oxide. O 2p orbitals are plotted in the left panels with the insets showing the zoom-in views from VBM to CBM, and corresponding data for Zr 4d are shown in right panels. The Fermi levels in the interface cells are shown by vertical dotted lines.

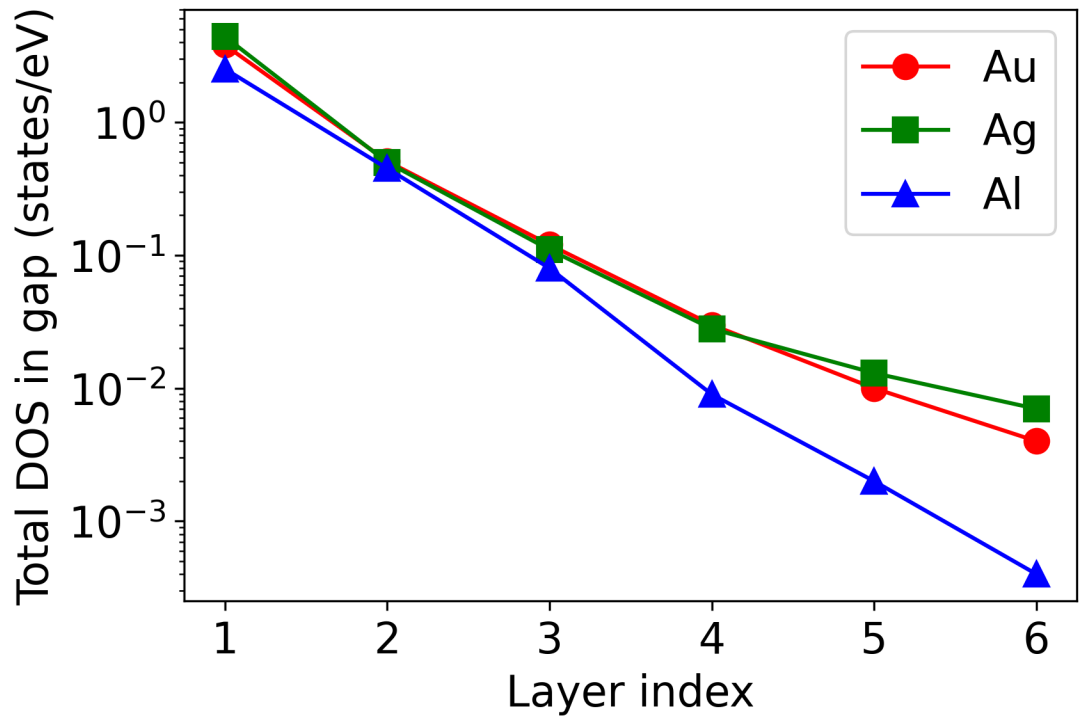

Figure 7. Decay of total MIGS within the gap region over the oxide layers. Note that the y-axis is in log scale.

The presence of MIGS leads to substantial deviations of the SBHs directly obtained from DFT, denoted $\Phi_B^{DFT}$, compared to those predicted by the ideal Schottky-Mott (SM) rule, $\Phi_B^{SM}$, which relates the SBH solely to the metal work function and the oxide electron affinity. Following the approach of Therrien et al. [79] $\Phi_B^{DFT}$ is determined as the energy difference between the CBM in the bulk-like oxide region and the Fermi level of the interface supercell, extracted from the layer-resolved local density of states (LDOS), as shown in Figures 8(a)-(c).

We note that the apparent curvature of the valence-band edges near the interface arises from the presence of MIGS rather than from electrostatic band bending. In intrinsic $tZrO_2$, the absence of free carriers precludes the formation of a space-charge region, and the limited size of the DFT supercell further prevents the resolution of long-range space-charge redistribution. Therefore, the SBHs extracted here reflect purely the short-range interfacial electronic effects, including MIGS formation, interface dipoles, and chemical bonding, without contributions from continuum-scale electrostatics.

As summarized in Figure 8(d) and Table II, $\Phi_B^{DFT}$ follows the trend Al-$tZrO_2$ < Ag-$tZrO_2$ < Au-$tZrO_2$, which is consistent with the ordering expected from the Schottky-Mott rule and underscores the important role of the metal work function. However, the absolute values of $\Phi_B^{DFT}$ are significantly lower than $\Phi_B^{SM}$, highlighting the critical role of MIGS in reducing the barrier height via Fermi-level pinning. This behavior is consistent with classical MIGS theory developed for metal-semiconductor and metal-insulator interfaces, where evanescent metal states introduce gap states that partially decouple the SBH from the metal work function [11,80].

Among the three systems, the Al-$tZrO_2$ interface exhibits the largest deviation from the Schottky-Mott limit. This pronounced deviation reflects substantial contributions from interfacial chemical bonding and strong Al-O hybridization, which alter the local electronic structure beyond what is captured by the metal-induced gap state (MIGS) picture alone. In contrast, the noble-metal interfaces are largely governed by MIGS-driven electronic effects with minimal chemical interaction, leading to more moderate deviations from the Schottky–Mott prediction. To quantify the role of MIGS at the noble-metal interfaces, we fitted the DFT-computed SBH ($\Phi_B^{DFT}$) for Ag-$tZrO_2$ and Au-$tZrO_2$ interfaces using Tersoff's MIGS-based model,

$$\Phi_B^{Tersoff} = S * \Phi_B^{SM} + \Phi_0, \quad (1).$$

where the pinning parameter $S$ characterizes the degree of Fermi-level pinning, with $S = 0$ corresponding to complete pinning and $S = 1$ to the Schottky-Mott limit. Using $\Phi_B^{DFT}$ and the work functions for Au and Ag, we obtain $S \approx 0.75$, indicating partial pinning at these interfaces. Assuming that the deviation of $\Phi_B^{DFT}$ from $\Phi_B^{SM}$ arises solely from MIGS, we extrapolate the fitted relation to the Al-$ZrO_2$ system. As shown in Fig. 8(d), the MIGS contribution accounts for only ~35% of the total deviation of $\Phi_B^{DFT}$ from $\Phi_B^{SM}$, with the remaining discrepancy primarily attributable to interfacial chemical bonding. Although this extrapolation is idealized for simplicity, it qualitatively corroborates the dominant role of chemical bonding in determining the Schottky barrier height at the Al-t$ZrO_2$ interface.

Further insight is provided by the LDOS in the metal regions adjacent to the interface. For the Au-t$ZrO_2$ and Ag-t$ZrO_2$ interfaces, substantial metal-derived DOS extends to energies near the oxide valence band maximum, consistent with strong MIGS penetration into the gap. In contrast, the Al-$ZrO_2$ interface exhibits negligible metal DOS near the oxide VBM, in agreement with the reduced MIGS density and faster decay observed in Figures 6 and 7. This reinforces the conclusion that, for Al-t$ZrO_2$, the interfacial charge transfer and SBH reduction are dominated by localized chemical bonding rather than MIGS-mediated Fermi-level pinning.

Finally, it should be emphasized that the SBHs reported here are intrinsic interfacial barriers obtained in the absence of space-charge redistribution in the oxide. The additional modification of SBHs due to long-range space-charge effects, which can be significant in lightly doped or defective oxides such as t$ZrO_2$, will be addressed explicitly in the following section.

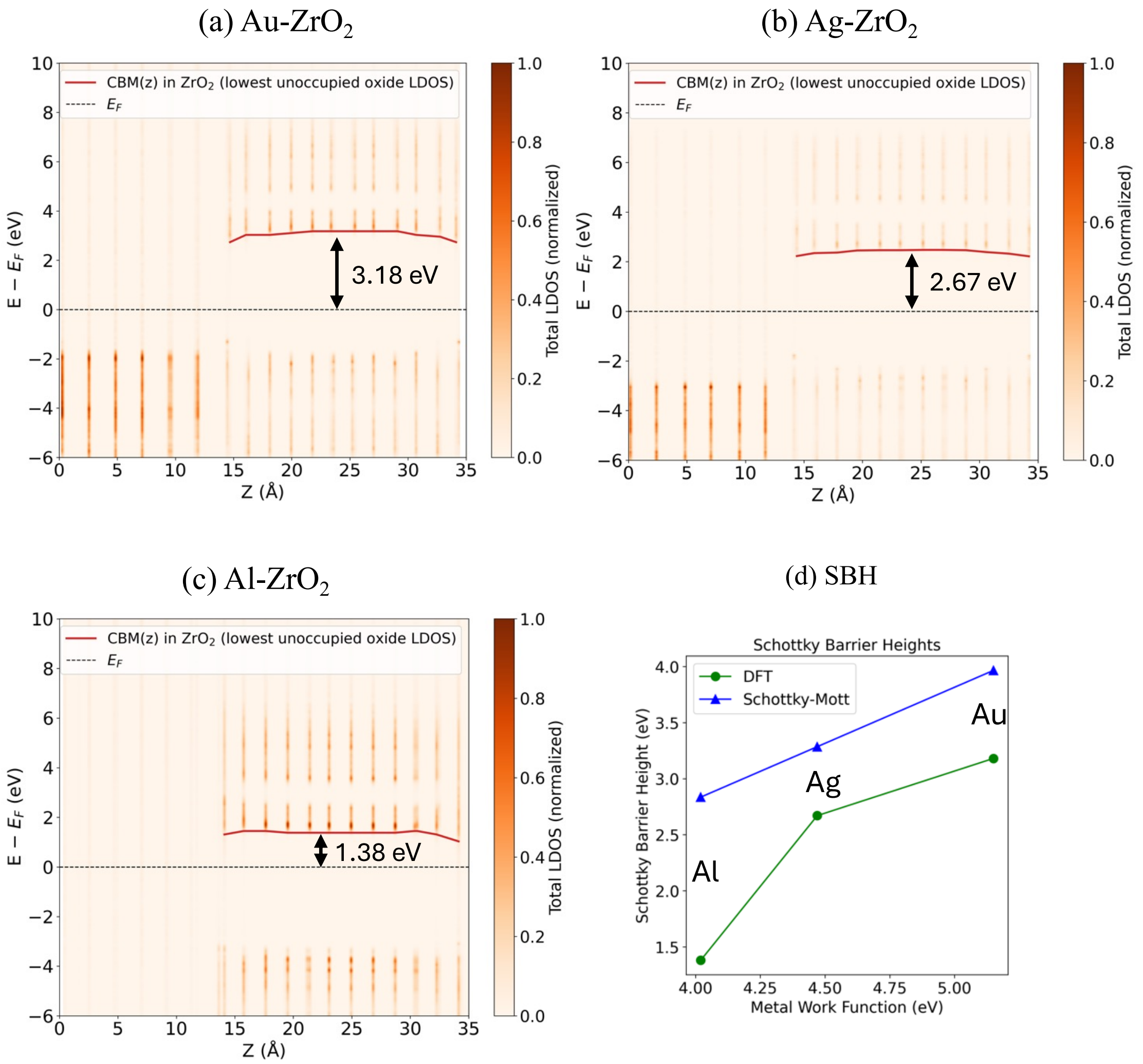

Figure 8. Normalized local density of states (LDOS) for (a) Au-t$ZrO_2$ (b) Ag-t$ZrO_2$ (c) Al-t$ZrO_2$ interfaces. The layer-resolved CBMs are shown by solid curves in reference to the Fermi energies (dotted lines). In (d), Schottky barrier heights (SBHs) obtained from DFT, calculated using Schottky-Mott's rule, and fitted following Tersoff's model are plotted against the metal work function.

### 3.4 Long-range space-charge redistribution

The preceding analysis based on explicit DFT calculations captures short-range charge redistribution confined to the immediate metal-oxide interface, typically within a few atomic layers. This regime is governed by MIGS, local chemical bonding, orbital hybridization, and the formation of an interfacial dipole. These effects directly modify the local electrostatic potential lineup and hence the intrinsic SBH at the atomic scale.

However, once the interface dipole is established and the Fermi levels of the metal and oxide align, the resulting electrostatic boundary condition necessarily induces a long-range redistribution of free carriers in the semiconductors or doped oxides. Consequently, while DFT defines the boundary conditions (interface dipole, SBH in the zero-field limit), the transport-relevant electrostatics are governed by continuum-scale band bending and space-charge formation. This redistribution manifests as a space-charge region and an associated potential drop that can extend over tens to hundreds of nanometers, far beyond the spatial reach accessible to DFT.

To capture this long-range response, we evaluate the electron DLW, surface charge accumulation, and effective SBH in n-type $tZrO_2$ using a continuum space-charge model. The electron concentration is varied from $5.86 \times 10^{-5}$ to $1.30 \times 10^{-1}$ nm$^{-3}$, spanning the dilute doping regime to the degenerate limit, which also includes several experimental doping concentrations [81–83]. The corresponding numerical results are summarized in Table S1 of the Supplementary Materials.

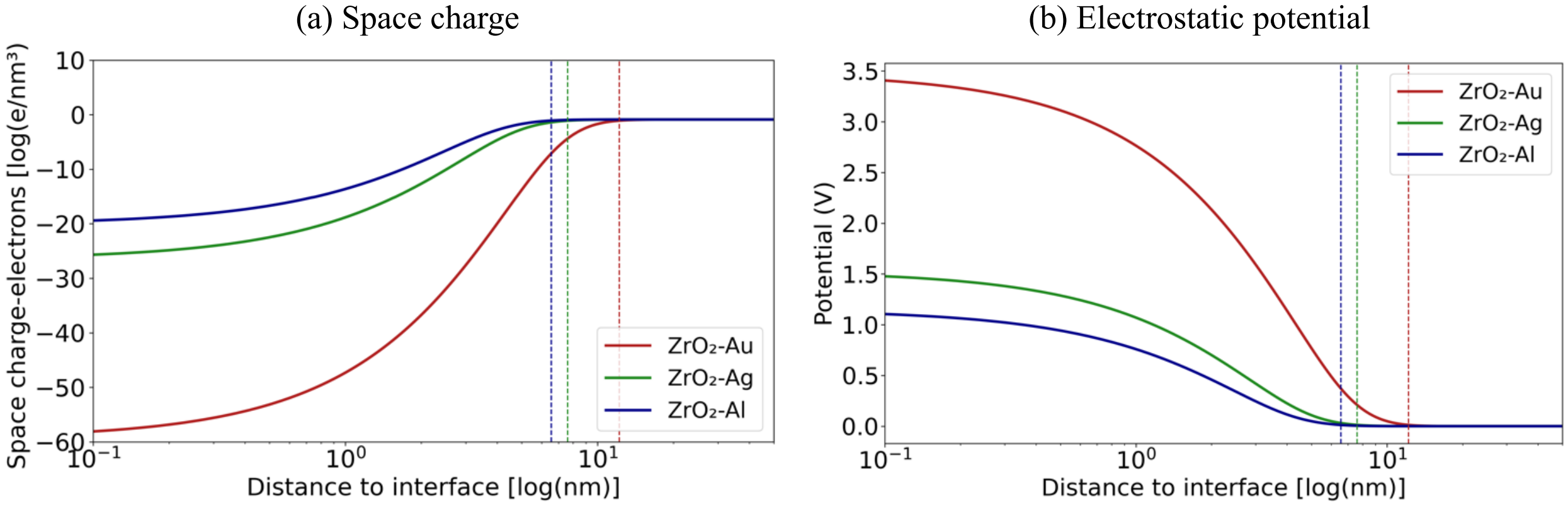

Figure 9. (a) Space-charge electron density and (b) electrostatic potential profiles as functions of the distance from the metal-oxide interface. The dotted vertical lines mark the boundaries of the depleted layer, at about 6.54, 7.59, and 12.18 nm for Al-$tZrO_2$, Ag-$tZrO_2$, Au-$tZrO_2$ interfaces, respectively. The doping concentration is $1.30\times10^{-1}$ e/nm$^{-3}$ in terms of bulk electron concentration and the temperature is 300 K temperature.

For all three metal contacts, Au, Ag, and Al, an electron depletion layer develops on the oxide side of the interface (Figure 9(a)), accompanied by a monotonic electrostatic potential drop (Figure 9(b)). The depletion region reflects the transfer of mobile electrons from the oxide toward the metal in response to Fermi-level alignment, leaving behind positively charged ionized donors in the oxide bulk. Charge neutrality is maintained by an equal and opposite surface charge accumulation localized near the interface. At the highest doping level considered ($1.30 \times 10^{-1}$ nm$^{-3}$), the DLW reaches 12.18 nm, 7.59 nm, and 6.54 nm for Au-t$ZrO_2$, Ag-t$ZrO_2$, and Al-t$ZrO_2$, respectively. These values exceed the charge-redistribution length scales obtained from DFT by roughly one order of magnitude, highlighting the intrinsic separation between atomic-scale interfacial physics and mesoscopic space-charge effects.

Importantly, both the magnitude of the surface charge accumulation and the total potential drop follow the systematic trend Au-t$ZrO_2$ > Ag-t$ZrO_2$ > Al-t$ZrO_2$, reflecting the decreasing metal work function. This demonstrates that the metal work function controls not only the short-range SBH but also the extent of long-range electrostatic perturbation in the oxide.

(a) Depletion layer width (DLW) (b) Schottky barrier height (SBH)

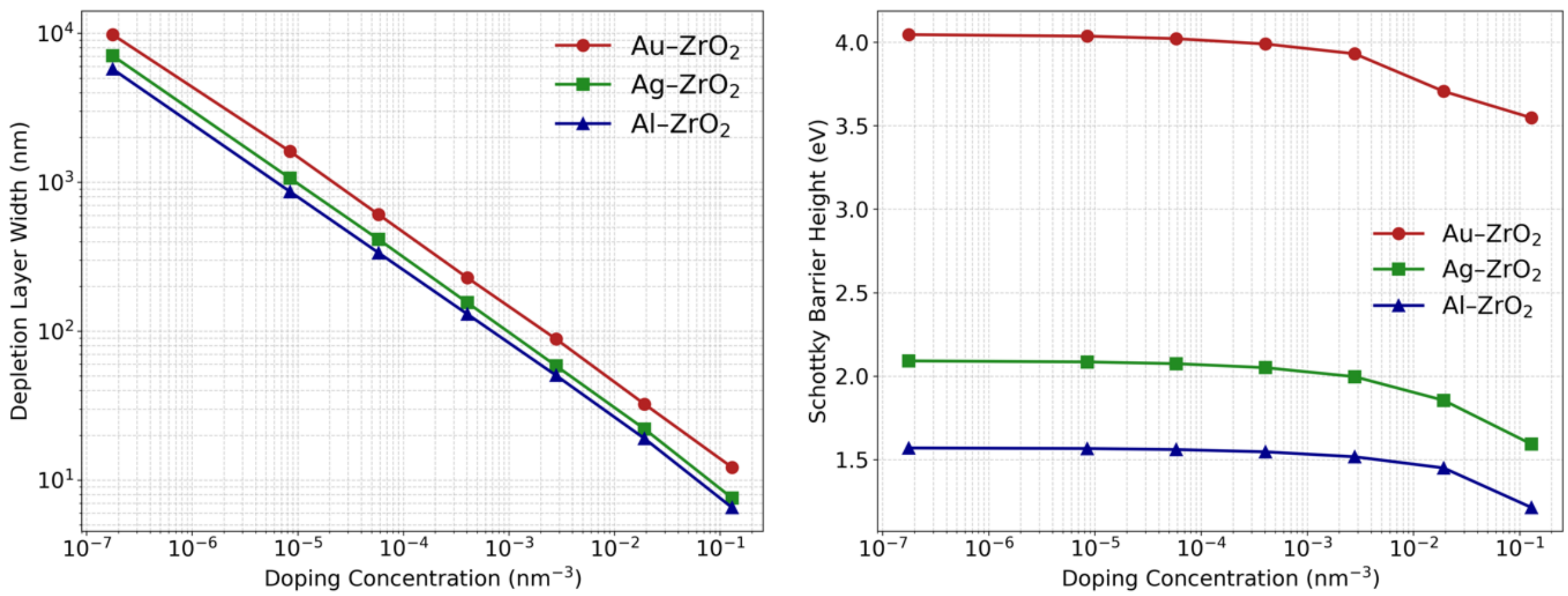

Figure 10. (a) Electron depletion layer width (DLW) and (b) Schottky barrier heights (SBH) versus doping concentration obtained from the Skachkov[24,69] model.

As shown in Figure 10(a), the DLW exhibits an approximately power-law dependence on the electron concentration, evidenced by the near-linear behavior in the log-log plot. Increasing doping reduces the DLW because a higher bulk carrier density can accommodate the required compensating charge over a shorter distance. This behavior is a direct consequence of Poisson's equation and is consistent with classical semiconductor junction theory.

Concurrently, the SBH decreases with increasing doping level (Figure 10(b)), with an increasingly steep slope at higher concentrations. This reduction arises from enhanced screening of the interface dipole and partial suppression of band bending in the degenerate regime. Both trends are in excellent qualitative agreement with the continuum analysis of Skachkov *et al.* [24] and with prior semiconductor junction models [84].

These results are also consistent with DFT studies by Jiao *et al.* [16] on the Au-$TiO_2$ interface, which demonstrate that doping - especially when localized near the interface - can substantially modify the SBH. It is important to note that, due to the limited size of DFT supercells, the effective doping levels accessible in DFT are comparable to the upper end of the doping range explored in the present continuum calculations, reinforcing the complementary nature of the two approaches. The SBH results from the continuum model are also in good agreement with various experimental SBH values for metal–$ZrO_2$ interfaces, such as 3.5 ± 0.1 eV for Au–$ZrO_2$ [85], 2.77 eV for Pt–$ZrO_2$ [86], and 0.92–1.06 eV for Al–$ZrO_2$ [87–89], further validating the model's accuracy.

Among the three interfaces, Au-t$ZrO_2$ exhibits the largest surface charge accumulation, followed by Ag-t$ZrO_2$ and Al-t$ZrO_2$, mirroring the trend in SBH and underscoring the dominant role of metal work function. At the highest doping level, the surface charge accumulation reaches approximately 1.13 e/nm$^2$, exceeding the net charge transfer of 0.45 e/nm$^2$ obtained from direct DFT calculations. This apparent discrepancy is not unexpected. In the continuum model, the surface charge arises solely from the redistribution of free carriers in response to band bending, whereas DFT captures localized charge rearrangements associated with MIGS, chemical bonding, and interfacial dipoles. As a result, neither the magnitude nor the sign of charge transfer should be directly compared between the two descriptions. Instead, they represent distinct but complementary contributions to the overall electrostatics: DFT defines the interfacial boundary condition within several angstroms to the interface, while the continuum model describes the long-range response of the doped oxide extend from nanometer scales up to macrometer scales from the interface, depending on the doping concentration in oxide.

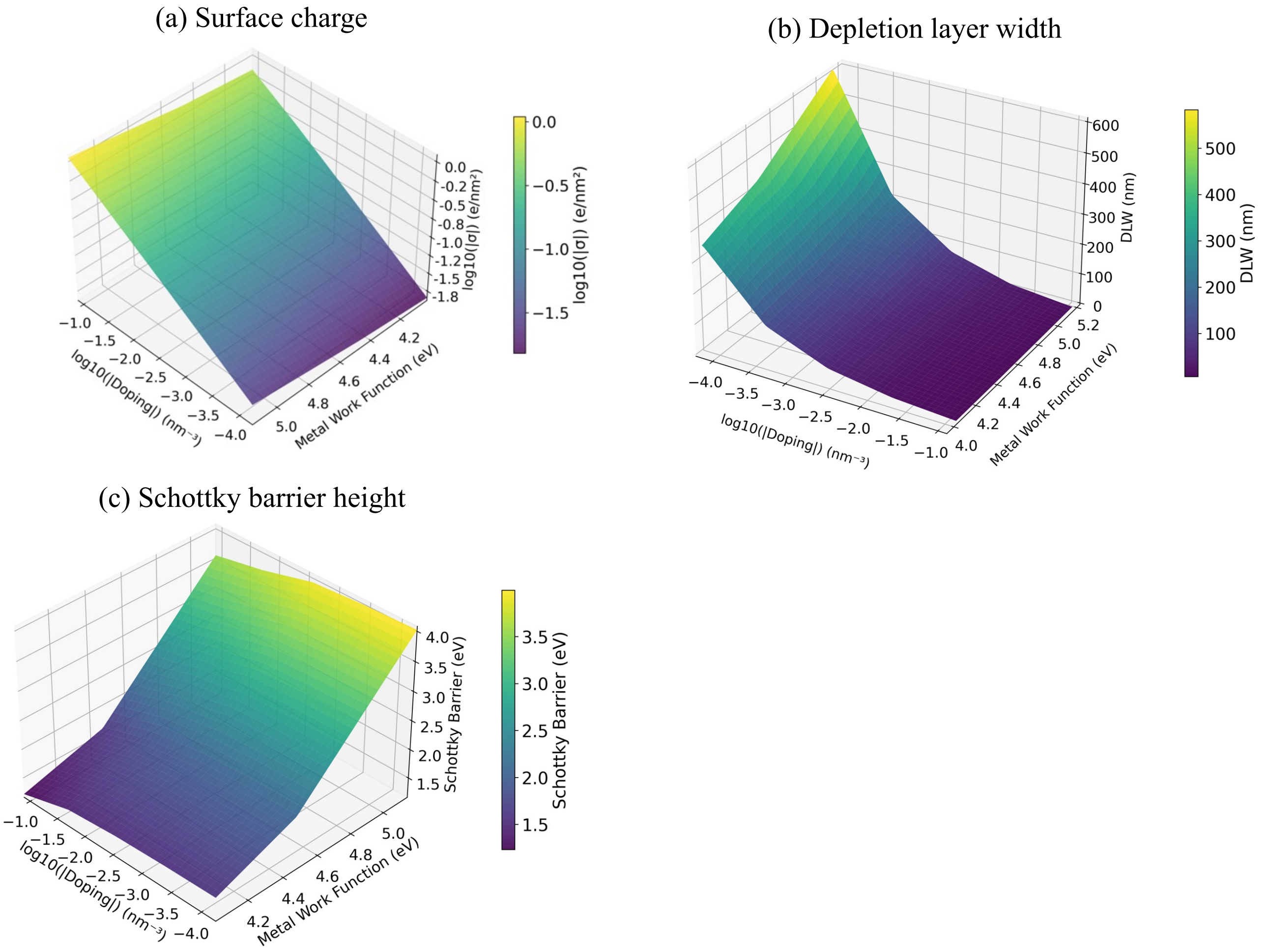

Figure 11. 3D plots of (a) surface charge, (b) depletion layer, and (c) Schottky barrier height, and width as functions of doping concentration and metal work function. For better visualization in (b), the doping concentration is set to increase from left side to right side

Figure 11 summarizes the dependence of surface charge accumulation, DLW, and SBH on both metal work function and doping level. All three quantities increase monotonically with metal work function, consistent with the increasing energy separation between the metal Fermi level and the oxide conduction band minimum. A higher work function thus produces a larger SBH, which in turn enhances electron depletion in the oxide and increases both the surface charge density and DLW. Notably, the surface charge and DLW are more sensitive to doping level than to metal work function (Figures 11(a) and 11(b)), reflecting their dependence on bulk carrier statistics. In contrast, the SBH depends more strongly on metal work function than on doping (Figure 11(c)), highlighting its origin in interfacial energy alignment rather than bulk electrostatics.

It should be noted that the results from the Skachkov's model [24] depend on CNL, which is defined as the energy level at which the integrated interfacial gap-state density of states is charge neutral. In practice, the CNL is extracted directly from the PDOS of the first metal and oxide layers. The resulting CNL depends on the interfacial properties, rather than a bulk property as originally assumed in Tersoff's model [80]. This procedure implicitly assumes that the interfacial gap states are dominated by MIGS and that contributions from interface-specific chemical bonding are negligible. In situations where these assumptions may not be fully satisfied, the extracted CNL becomes strongly system dependent, leading to significant uncertainties in the predicted electrostatic response. Also, the long-range electrostatic potential associated with space-charge redistribution is not accessible within the finite-size DFT framework. Consequently, while the present theoretical analysis provides a valuable framework for quantitatively linking atomistic interfacial physics obtained from DFT to long-range space-charge redistribution, a direct quantitative comparison between DFT-derived quantities and their continuum-model counterparts, such as SBH and surface charge density, is not appropriate.

To place the above considerations into a broader context, it is important to emphasize that the concept of CNL has proven highly valuable for establishing a physically transparent link between interface electronic structure and macroscopic Schottky barrier formation. In this regard, the present approach should be viewed as complementary to the original bulk-based formulation of Tersoff [80], rather than as a replacement. Within the current framework, the CNL is interpreted as an effective interfacial quantity, reflecting the electronic structure of the metal–semiconductor junction as obtained from atomistic DFT calculations. This interpretation naturally accounts for the fact that the electronic properties of real interfaces are strongly influenced by the microscopic state of the semiconductor surface. In particular, the interfacial CNL is expected to depend on surface preparation conditions, including chemical treatments, surface reconstructions, and the presence of charged defects or residual adsorbates, all of which modify the spectrum and occupancy of interfacial gap states. Consequently, adopting a CNL derived solely from bulk electronic properties may not fully capture the physics of realistic metal-semiconductor contacts, especially in cases where interface-specific effects play a significant role. At the same time, it is recognized that standard DFT calculations are intrinsically limited to a finite number of atomic layers and therefore cannot explicitly describe long-range electrostatic effects associated with space-charge redistribution in the semiconductor. From this perspective, the combination of atomistic DFT with a continuum electrostatic model offers a balanced and flexible description. While DFT provides a detailed microscopic picture of the interface electronic structure, additional physically motivated parameters in the SB model, such as surface or interface charge densities, allow one to incorporate effects beyond the finite-size DFT framework. This hybrid strategy enables a consistent connection between first-principles interface physics and experimentally relevant electrostatic responses, while maintaining transparency regarding the underlying assumptions and their domain of applicability.

The DFT calculations and the continuum electrostatic analysis are obtained with ideal, defect-free $ZrO_2$, to provide a foundational framework for evaluating oxide films formed on realistic Zr-cladding alloys and under realistic nuclear reactor operation conditions. $ZrO_2$ is a wide-bandgap oxide. While forming on pure Zr, the oxygen chemical potential at the metal-oxide interface is set by the Zr/$ZrO_2$ equilibrium in accordance with an exceptionally low oxygen partial pressure, $p_{O_2}$. In such an limit, the concentrations of intrinsic point defects, thereby the intrinsic carrier densities, are vanishingly small. This picture is consistent with the very small nonstoichiometry reported for $ZrO_2$ and with the strong dependence of its conductivity on oxygen-pressure [90]. In this case, the corresponding range of space-charge redistribution is described by the low-doping limit in the electrostatic analysis in Figure 10(a). However, trace impurities (*e.g.*, ppm-level aliovalent dopants or unintentional contaminants) are inevitable in practical materials processing, and they are sufficient to make $ZrO_2$ effectively doped and make carrier concentrations controlled by extrinsic defects rather than intrinsic disorder [91]. Considering that metallic Zr is extremely difficult to purify to the ppm level (particularly with respect to residual aliovalent impurities) [92], even thermally grown $ZrO_2$ formed on nominally “pure” Zr should be regarded as extrinsically doped. Moreover, albeit the limited solubility of the deliberate alloying additions used in nuclear fuel cladding (such as Sn, Ni, Nb, Fe, Cr), their bulk concentrations are at least on the order of several hundred ppm, implying that the corrosion oxides formed during reactor operation are unambiguously doped, with defect populations and carrier densities controlled primarily by these extrinsic species. In such a case, the high-doping limit in Figure 10(a) is appropriate for estimating the range of space-charge redistribution.

Furthermore, $ZrO_2$ formed during reactor operation is subject to neutron irradiation. As such, lattice defects such as oxygen vacancies [93–95] will be produced, and their concentrations can reach orders of magnitude higher than their thermal equilibrinium values. Oxygen vacancies are expected to act as donor-like centers, introducing excess electrons and shifting the Fermi level toward the conduction band of $ZrO_2$, while oxygen interstitials and Zr vacancies are expected to act as acceptor-like. Within this framework, the Schottky barrier heights and charge redistribution profiles reported here for the Au-$ZrO_2$, Ag-$ZrO_2$, and Al-$ZrO_2$ interfaces should be regarded as an idealized limit, against which defect-induced modifications can be qualitatively assessed. The differences observed among the three metal-$ZrO_2$ systems indicate that the influence of lattice defects on interfacial electronic properties can be strongly metal-dependent. For metals with higher chemical reactivity toward oxygen, such as Al, the stronger interfacial bonding and more pronounced charge redistribution are likely to enhance the coupling between vacancy-induced donor states and metal-induced gap states, promoting Fermi-level pinning and local barrier thinning. In contrast, more inert metals such as Au are expected to exhibit weaker coupling between vacancy states and the metal electronic structure, thereby preserving a comparatively more stable Schottky barrier even in the presence of moderate defect concentrations. These trends underscore the critical role of the intrinsic electronic structure of the ideal metal–$ZrO_2$ interface for understanding how defect populations in $ZrO_2$ translate into macroscopic charge transport behavior and, ultimately, corrosion kinetics, which is an area vastly open.

## 4. Conclusion

In this work, we present a unified, multiscale description of charge redistribution at metal-t$ZrO_2$ interfaces by explicitly connecting atomic-scale interfacial physics with long-range space-charge electrostatics. DFT calculations reveal pronounced short-range charge redistribution confined to a few atomic layers near the interface, governed by metal-induced gap states, interfacial dipoles, and chemical bonding. The MIGS arises primarily from the O 2p orbital near VBM and from Zr 4d near CBM. For noble metals like Ag and Au, the short-range charge distribution is dominated by MIGS and the resulting interfacial dipole. While for active metals like Al, the importance of chemical bonding dominates over MIGS and interfacial dipole, leading to opposite electron transfer direction to that predicted by metal work function. Additionally, the difference in the coupling behavior of Al 3s and 3p with O 2p than that of Au 5d and Ag 4d gives weaker and fast-decaying MIGS. In addition to these atomic-scale insights obtained with intrinsic tZrO2 without free carriers, these effects define the intrinsic SBH and establish the electrostatic boundary conditions for the adjoining oxide.

Building upon these interfacial physics from atomic scale, continuum-scale modeling captures the long-range response of doped $tZrO_2$ to Fermi-level alignment. For all metal contacts studied, electron depletion layers develop on the oxide side, accompanied by extended potential drops that can span orders of magnitude larger than the spatial extent of charge redistribution accessible to DFT, particularly at low doping levels. The magnitude of the surface charge accumulation, DLW, and SBH correlate positively with the metal work function, with Au-$tZrO_2$ > Ag-$tZrO_2$ > Al-$tZrO_2$, highlighting its dominant role in controlling long-range electrostatics.

We further demonstrate the importance of doping in modulating the long-range response. Increasing electron concentration significantly reduces the DLW and partially suppresses the SBH through enhanced electrostatic screening. While the SBH is primarily controlled by metal work function, the surface charge accumulation and DLW exhibit stronger sensitivity to doping level, consistent with classical space-charge theory and prior continuum models.

Overall, this study clarifies the complementary roles of short-range and long-range charge redistributions in determining the electrostatics of metal-oxide interfaces by explicitly bridging DFT and continuum descriptions. Our results provide a consistent framework for interpreting Schottky barrier formation, carrier transport, and defect behavior across length scales. Importantly, these results also establish a mechanistic basis for understanding how NMIs embedded in $ZrO_2$ or corrosion oxides can modulate charged-species transport via space-charge effects, which can extend far beyond their interfaces. As shown here, each NMI acts as an internal metal-oxide junction that imposes a work-function-dependent electrostatic boundary condition, generating a local space-charge region and built-in field whose spatial extent is governed by oxide doping. These local fields can bias the migration of charged defects and carriers (such as oxygen vacancies and electrons), producing heterogeneous fluxes and thereby promoting spatially non-uniform oxide growth in the vicinity of inclusions. The space charge effects on charge transport will be the subject of future work. This multiscale perspective is broadly applicable to oxide electronics, resistive switching materials, and metal-oxide contacts where both interfacial chemistry and bulk doping play critical roles.

**CRediT authorship contribution statement**

**Ximeng Wang**: running calculations, writing, review & editing. **Dmitry Skachkov**: running calculations, revieiw & editing. **Arnab Das**: Review. **Junliang Liu**: Review. **Alexander Kvit**: Review. **Jennifer Choy**: Review & editing. **Adrien Couet**: Review & editing. **Yongfeng Zhang**: Supervision, writing, review & editing.

**Declaration of competing interest**

The authors declare that they have no known competing financial interests or personal relationships that could have appeared to influence the work reported in this manuscript.

**Acknowledgements**

This work was supported by the U.S. Department of Energy, Office of Science, Basic Energy Sciences under Award DE-SC0023873. This research made use of the resources of the High-Performance Computing Center (HPC) at Idaho National Laboratory, which is supported by the Office of Nuclear Energy of the U.S. Department of Energy and the Nuclear Science User Facilities. We also thank Dr. Xiaoguang Zhang from the University of Florida and Dr. Anuj Goyal from the Indian Institute of Technology Hyderabad for their valuable comments.